\documentclass[preprint,showpacs,aps]{revtex4}
\usepackage{epsfig}
\usepackage{graphicx}
\usepackage{amsmath}
\usepackage{amsfonts,amssymb,amsmath,amsthm}

\newcommand{\be}{\begin{equation}}
\newcommand{\ee}{\end{equation}}
\newcommand{\ba}{\begin{eqnarray}}
\newcommand{\ea}{\end{eqnarray}}

\begin{document}

\title{Abelian Chern-Simons term as a Kaluza-Klein dimensional reduction of the Gibbons-Hawking surface term}

\author{Hongsu Kim\footnote{e-mail: chris@kasi.re.kr}}

\affiliation{Department of Astronomy and Space Science,   College of Natural Sciences, Chungnam National University, Daejeon
34134, Korea}

\author{Jae Sok Oh\footnote{e-mail: ojs001@kasi.re.kr}}

\affiliation{Korea Astronomy and Space Science Institute, Daejeon 34055, Korea}

\begin{abstract}
It is suggested that the original, minimal Kaluza-Klein theory should be extended by adding a 5-dimensional version of the Gibbons-Hawking gravitational surface term.
It is then demonstrated that the usual dimensional reduction of the newly added surface (boundary) term leads to the emergence of the famous Abelian Chern-Simons term.
It is stressed that the advent of this Chern-Simons term is not merely a parametrization artefact but a real thing. Finally, the issue of finite-ranged electromagnetic 
interaction due to massive photons on a plane has been interpreted in terms of the violation of the local gauge invariance of this extended version of Kaluza-Klein theory.   
\end{abstract}



\maketitle


\newpage
\begin{center}
{\rm\bf I. INTRODUCTION}
\end{center}

The suggestion of Kaluza \cite{Kaluza} and Klein \cite{Klein} that electromagnetism and general relativity could be unified by starting with a 5-dimensional version of pure gravity has been one of the most intriguing and elegant way of unifying gauge theories with gravitation. According to their hypothesis, the world has (4+1)-spacetime dimensions, but the additional spatial dimension is 'curled up' to form a circle so small as to be unobservable. They, then, showed that ordinary general relativity in 5-dimensions, assuming such a cylindrical ground state, contained a local U(1) gauge symmetry arising from the isometry of the hidden 5th dimension. At this point, let us turn our attention to the complete form of the action for pure gravity in general. The action for pure gravity is usually taken to be of the Einstein-Hilbert form, $\frac{1}{2\kappa^2} \int d^{4}x \sqrt{g} R$ in 4-dimensions in which the scalar curvature contains forms linear in second derivatives of the metric. As was pointed by Gibbons and Hawking \cite{GH}, however, in order to obtain an action which depends on the first derivatives of the metric, as is required by the path integral approach in quantum gravity, the second derivatives have to be removed by integration by parts.

The complete action, for the metric $ {g}$ over a manifold $ {M}$ with boundary $\partial {M}$, then, has the form 
\begin{eqnarray}
   S = \frac{1}{2\kappa^2} \int_{M} d^{4}x \sqrt{g} R 
       + \frac{1}{\kappa^2}  \int_{\partial M} d^{3}x \sqrt{g_{3}} K                      
    \nonumber  
\end{eqnarray}
with ${K}$ in the surface term being the trace of the second fundamental form i.e., extrinsic curvature of the boundary $\partial {M}$ in the metric ${g}$. Here the surface term on $\partial {M}$ has been chosen so that for metric $ {g}$ which satisfy the Einstein equations, the action is an extremum under variation of the metric which vanishes on the boundary $\partial {M}$ but which may have non-zero normal derivatives. Note also that the addition of the surface term above is known to be needed in the ADM's space-plus-time split formalism for the canonical formulation of general relativity \cite{ADM}. 

Therefore we propose here that when considering the 5-dimensional pure gravity as the starting point of the Kaluza-Klein unification, the complete gravity action should involve a 5-dimensional version of the Gibbons-Hawking surface term described above. The consequence of the addition of this surface term is the emergence of another magic over that of Kaluza and Klein; the natural appearance of the Abelian Chern-Simons term on the 3-dimensional boundary as a result of the dimensional reduction of the newly added Gibbons-Hawking surface term. The crucial element in obtaining this conclusion is that the scalar dilaton field $\phi (x)$, namely the 15th degree of freedom in the 5-dimensional spacetime metric $ \hat {g}_{55}$ should be allowed to vary in contrast to Kaluza and Klein who simply had set it equal to one.    

To demonstrate this explicitly, one begins with the complete 5-dimensional pure gravity action with the surface term:
\begin{eqnarray}
   \hat{S} = \frac{1}{2\hat{\kappa}^2} \left[ \int_{M} d^{5}\hat{x} \sqrt{\hat{g}} \hat{R} 
       + 2 \int_{\partial {M}} d^{4}\hat{x} \sqrt{\hat{g}_{4}} \hat{K} \right]                       
\end{eqnarray}
where $\hat{\kappa}^2 = 8\pi G_{5}$ is a 5-dimensional counter part of the gravitational constant.

\begin{center}
{\rm\bf II. Kaluza-Klein parametrization 1 for the dimensional reduction}
\end{center}

Following Kaluza and Klein, we perform the dimensional reduction via the (4+1) split 
\begin{eqnarray}
   \hat{g}_{\hat{\mu}\hat{\nu}} = e^{\frac{\phi}{\sqrt{3}}}
   \begin{pmatrix}
   g_{{\mu}{\nu}} + e^{-\sqrt{3}\phi} A_{\mu}A_{\nu}   &  e^{-\sqrt{3}\phi} A_{\mu} \\
   e^{-\sqrt{3}\phi} A_{\nu}                           &  e^{-\sqrt{3}\phi}         \\
   \end{pmatrix}
\end{eqnarray}
In equations (1) and (2), $\hat{\mu}, \hat{\nu} = 0,1,2,3,4$ and all hatted quantities are 5-dimensional and unhatted 4-dimensional. Thus, by making this (4+1) split of the 5-dimensional Einstein equations, $\hat{R}_{\hat{\mu}\hat{\nu}} = 0$, Kaluza miraculously showed that $g_{\mu\nu}(x,y)$, $A_{\mu}(x,y)$ and $\phi(x,y)$ are spin-2 graviton, spin-1 photon, and spin-0 scalar dilaton since they satisfy the Einstein equations, the Maxwell equation, and the massless Klein-Gordon equation, respectively. Attractive though his idea was, Kaluza failed to provide a rationale for the suppression of the dependence on the extra coordinate $y$ and its manifest invisibility. To these problems, Klein supplied a resolution by suggesting that one should assume the 5th dimension as having circular ($S^1$) topology so that the coordinate $y$ is periodic $0 \le y/R \le 2\pi$ with $R$ being the tiny circumference of the compact 5th dimension.  

Then the 5-dimensional manifold in equation (1) has the topology of $M = M_{4} \times S^{1}$ and all fields quantities are periodic functions of the coordinate $y$ on the circle, i.e., any field quantity $F(x, y)$ (with $F$ being any of the $g_{\mu\nu}(x,y)$'s, $A_{\mu}(x,y)$'s and $\phi(x,y)$) admits a Fourier expansion;
\begin{eqnarray}
   F(x, y) = \sum_{n=-\infty}^{\infty} F_{n}(x) e^{iny/R}                 
\end{eqnarray}
so that the Kaluza-Klein theory describes an infinite number of 4-dimensional fields.  

Klein then assumed the "cylindrical ground state" condition: He truncated the action 
\begin{eqnarray}
   \hat{S} = \frac{1}{2\hat{\kappa}^2} \int_{M} d^{5}\hat{x} \sqrt{\hat{g}} \hat{R} 
   \nonumber                      
\end{eqnarray}
by dropping all harmonics with $n \neq 0$, retaining only the zero modes to obtain the dimensionally reduced 4-dimensional action of the form:   
\begin{eqnarray}
   S = \frac{1}{2\kappa^2} \int_{\bf{M_4}} d^{4}x \sqrt{g} 
       \left[ R - \frac{1}{2}\nabla_{\mu}\phi \nabla^{\mu}\phi - \frac{1}{4}e^{-\sqrt{3}\phi}F_{\mu\nu}F^{\mu\nu} \right]                       
\end{eqnarray}
where $2 \kappa^2 = 2\hat{\kappa}^2 (2\pi R)^{-1}, g = - \det g_{\mu\nu}$ and $F_{\mu\nu} = \nabla_{\mu}A_{\nu} - \nabla_{\nu}A_{\mu} $ with all field quantities depending only on $x^{\mu}$.

As is well-known, the reduced 4-dimensional action in equation (4) exhibits invariance under the 4-dimensional general coordinate transformations, the local U(1) gauge transformations, and the global scale transformations. And among other things, the reduced action in equation (4) describes the Einstein-Maxwell-Dilaton theory confirming Kaluza's discovery.

It is, however, puzzling that in all early works, by Kaluza and by Klein, the scalar dilaton field $\phi (x)$ is always set equal to one without much worry about the constraint it leads to; $F_{\mu\nu}F^{\mu\nu} = 0$. The realization of the importance of including the scalar field $\phi(x)$ was not pointed out until 1940's when Jordan \cite{JO} and later Thirty \cite{TH} considered the full action (4).  

Now, we turn to the new ingredient we attempt to include into this Kaluza-Klein theory. Since we now have the Gibbons-Hawking surface term in the 5-dimensional pure gravity action in equation (1), we also need to perform the dimensional reduction of it via the same type of spirit as in equation (2). To do so, we begin by taking the boundary term as being defined on a, say, constant $x^{i}$ hypersurface ($i \neq 0, 4$) so that the 4-dimensional boundary has the topology of $\partial M = \partial M_{4} \times S^{1}$. Then the extrinsic curvature of this hypersurface is given by
\begin{eqnarray}
   \hat{K}_{\hat{a}\hat{b}} = \nabla_{\hat{a}}n_{\hat{b}}, \hat{K} = \hat{g}^{\hat{a}\hat{b}} \hat{K}_{\hat{a}\hat{b}} 
   \nonumber                      
\end{eqnarray}
where $n^{\hat{a}}$ is the unit tangent to the congurence of geodesics orthogonal to $\partial M$.

Using $\hat{K} = \nabla_{\hat{a}} n^{\hat{a}} = \frac{1}{\sqrt{\hat{g_4}}} \partial_{\hat{a}} \left[ \sqrt{\hat{g_4}} n^{\hat{a}} \right]$, now we have,  
\begin{eqnarray}
   \int_{\partial M_{4} \times S^{1}} d^{4}\hat{x} \sqrt{\hat{g_4}} \hat{K} 
   = \int_{\partial M_{4} \times S^{1}} d^{4}\hat{x} \left[ n^{\hat{a}} \partial_{\hat{a}} (\sqrt{\hat{g_4}}) + \sqrt{\hat{g_4}} (\partial_{\hat{a}} n^{\hat{a}}) \right]                       
\end{eqnarray}
where $\hat{g}_{\hat{a}\hat{b}}$ is the metric ("pullback") on the 4-dimensional surface, $x^{i} = const.$, thus $\hat{a}, \hat{b} = 0, 2, 3, 4$.  
 
For our choice of the parametrization of the metric in equation (2), on the surface $\partial{M} = \partial M_{4} \times S^{1}$, $\hat{g_4} = -\det \hat{g_{4}}_{{\hat{a}}{\hat{b}}} = e^{\frac{\phi}{\sqrt{3}}} g_{3}$ with $g_{3ab}$ being the metric on $\partial {M_4}$.  

Therefore, using the familiar formula, $\Gamma^{\hat{b}}_{\hat{b}\hat{a}} = \frac{1}{\sqrt{\hat{g_4}}} \partial_{\hat{a}} (\sqrt{\hat{g_4}}) = \frac{1}{2} \partial_{\hat{a}} (\ln \hat{g_4})$, we have
\begin{eqnarray}
   \int_{\partial M_{4} \times S^{1}} d^{4}\hat{x} \sqrt{\hat{g_4}} \hat{K} 
   &=& \int_{\partial M_{4} \times S^{1}} d^{4}\hat{x} \sqrt{g_3} e^{{\phi}/2\sqrt{3}} \left[ \frac{1}{2\sqrt{3}} n^{\hat{a}} (\partial_{\hat{a}} \phi) + \partial_{\hat{a}} n^{\hat{a}} + \frac{1}{2} n^{\hat{a}} \partial_{\hat{a}} (\ln g_{3}) \right] \nonumber \\ 
   &=& 2\pi R \int_{\partial{M_4}} d^3x\sqrt{g_3} e^{{\phi}/2\sqrt{3}} \left[ \frac{1}{2\sqrt{3}} n^{a} (\partial_{a} \phi) + K \right]                      
\end{eqnarray}
where in the last line, we likewise assumed cylindrical ground state condition to carry out the $y$ integration and identified $K \equiv \nabla_a n^a = (\partial_{a} n^{a} + \Gamma^{b}_{ba})$ with the extrinsic curvature of the 3-dimensional boundary $\partial{M_4}$. The surface term on $\partial{M_4}$ given in equation (6) obtained from the dimensional reduction of the surface term on $\partial M$ does not, however, seem to admit immediate, clear physical interpretation in the present form, yet. Naturally, therefore, we seek for its "on-shell" equivalent (i.e., classically equivalent) action related possibly to it via the classical field equations. The classical field equations obtained by extremizing the dimensionally reduced action given in equation (4), of course, remains unchanged upon adding the surface term (6) in the action (one can easily confirm this through the straightforward calculation).

The classical field equations for photon, dilaton, and metric are given, respectively, by         
\begin{eqnarray}
   \nabla_{\mu} (e^{-\sqrt{3}\phi} F^{\mu\nu}) = 0, \nonumber \\
   \nabla_{\mu} \nabla^{\mu} \phi + \frac{\sqrt{3}}{4} e^{-\sqrt{3}\phi} F_{\mu\nu}F^{\mu\nu} = 0, \\ 
   R_{\mu\nu} = \frac{1}{2} \left[ \nabla_{\mu}\phi \nabla_{\nu}\phi 
                + e^{-\sqrt{3}\phi} (F_{\mu\lambda} F^{\lambda}_{\nu} - \frac{1}{4} g_{\mu\nu} F_{\alpha\beta} F^{\alpha\beta}) \right]. \nonumber  
\end{eqnarray}

With the help of the photon equation of motion, one can integrate the dilaton equation of motion once to yield
\begin{eqnarray}
   \sqrt{g} ( \nabla^{\mu}\phi + \frac{\sqrt{3}}{2} e^{-\sqrt{3}\phi} A_{\nu} F^{\mu\nu}) = const. \nonumber  
\end{eqnarray}
Thus on the 3-dimensional boundary $\partial_{M_4}$, we have, at the classical level,
\begin{eqnarray}
   \nabla_{a} \phi + \frac{\sqrt{3}}{2} e^{-\sqrt{3}\phi} A^{b} F_{ab} = 0  
\end{eqnarray}
where we chose the integration constant to be zero.

Now by making use of equation (8), we can rewrite the dimensionally-reduced surface term (6) into its on-shell equivalent action of the form:
\begin{eqnarray}
   \int_{\partial M_{4} \times S^{1}} d^{4}\hat{x} \sqrt{\hat{g_4}} \hat{K} 
    = 2\pi R \int_{\partial{M_4}} d^3x \sqrt{g_3} e^{{\phi}/2\sqrt{3}} \left[ K - \frac{1}{4} e^{-\sqrt{3}\phi} n^{a} A^{b} F_{ab} \right]                      
\end{eqnarray}
The second term on the right hand side of equation (9) may be rewritten using the antisymmetry of the photon field sterngth tensor under the interchange of its indices, $F_{ab} = - F_{ba}$, namely, 
\begin{eqnarray}
   n^a A^b F_{ab} = \frac{1}{2} (n^a A^b - n^b A^a)F_{ab} = \frac{1}{2} \epsilon^{abc} A_{c} F_{ab} = \epsilon^{abc} A_c \partial_a A_b, \nonumber  
\end{eqnarray} 
where we used that $n^a$ is just a unit vector orthogonal to the 3-dimensional boundary $\partial{M_4}$. Notice that this term, therefore, can be identified with the Abelian Chern-Simons term defined on a 3-dimensional manifold. As promised, we have shown explicitly that the inclusion of the Gibbons-Hawking surface term, upon dimensional reduction, leads to the emergence of the Abelian Chern-Simons term on the 3-dimensional boundary $\partial{M_4}$ of the 4-dimensional manifold $M_4$ in which the usual Maxwell term resides. 

To summarize, the complete 5-dimensional pure gravity action with the surface term in equation (1), reduces, upon the Kaluza-Klein dimensional reduction, to the following action for Einstein-Maxwell-Chern-Simons theory with dilaton defined on a 4-dimensional manifold and its 3-dimensional boundary,
\begin{eqnarray}
   S &=& \frac{1}{2\kappa^2} \int_{M_4} d^{4}x \sqrt{g} R + \frac{1}{\kappa^2} \int_{\partial{M_4}} d^{3}x \sqrt{g_3} e^{\frac{\phi}{2\sqrt{3}}} K  \nonumber  \\  
     &+& \frac{1}{2\kappa^2} \int_{M_4} d^{4}x \sqrt{g} \left(-\frac{1}{2} \nabla_{\mu}\phi \nabla^{\mu}\phi - \frac{1}{4} e^{-\sqrt{3}\phi} F_{\mu\nu}F^{\mu\nu}\right)  \\
     &+& \frac{1}{2\kappa^2} \int_{\partial{M_4}} d^{3}x \sqrt{g_3} e^{\frac{\phi}{2\sqrt{3}}} \left(-\frac{1}{4} e^{-\sqrt{3}\phi} \epsilon^{abc}A_{a}F_{bc}\right)  
     \nonumber                     
\end{eqnarray}

Now we would like to stress that the emegence of the Abelian Chern-Simons term as a result of the Kaluza-Klein dimensional reduction of the Gibbons-Hawking surface term is not merely an artefact or a coincidence but indeed a general, generic feature. To see this, it might worth noting that the conformal factor, $e^{\phi/\sqrt{3}}$ in the choice of the parametrization of the 5-dimensional metric given in equation (2) is to ensure that upon dimensional reduction, no non-minimal coupling term appears, i.e., no extra factors of $e^{-\sqrt{3}\phi}$ multiply the scalar curvature $R$ in the (minimally) dimensionally-reduced 4-dimensional action (4). Thus the first term in the reduced action (4) looks precisely like the ordinary Einstein-Hilbert action. Therefore if we had chosen some other parametrization of the 5-dimensional metric, the dimensionally-reduced action $S$ would have looked differently possbly involving a non-minimal coupling term. Still, however, all physical predictions would remain the same. 

\begin{center}
{\rm\bf III. Kaluza-Klein parametrization 2 for the dimensional reduction}
\end{center}

We, therefore, expect that generally the emergence of the Chern-Simons term from the Gibbons-Haeking surface term would happen regardless of the different choice of the parametrization of the 5-dimensional metric. Indeed, this can be readily checked, too. For instance, with the choice of the (different) parametrization,    
\begin{eqnarray}
   \hat{g}_{\hat{\mu}\hat{\nu}} = 
   \begin{pmatrix}
   g_{{\mu}{\nu}} + \phi A_{\mu}A_{\nu}     &    \phi A_{\mu} \\
   \phi A_{\nu}                             &    \phi         \\
   \end{pmatrix} 
\end{eqnarray}
the dimensionall-reduced 4-dimensional action takes the form;
\begin{eqnarray}
   S = \frac{1}{2\kappa^2} \int_{M_4} d^{4}x \sqrt{g} \phi^{1/2} \left[ R - \frac{2}{\phi^{1/2}} \nabla_{\mu}\nabla^{\mu} \phi^{1/2} 
                                                                        - \frac{1}{4} \phi F_{\mu\nu} F^{\mu\nu} \right]  
\end{eqnarray}
where the dilaton action term $\sim 2\partial_{\mu} (\sqrt{g} g^{\mu\nu} \partial_{\nu} \phi^{1/2})$ is "to drop out" since it is a total derivative. And the classical field equations for photon, dilaton, and metric obtained by extremizing above action are given, respectively, by:
\begin{eqnarray}
   \nabla_{\mu} (\phi^{3/2} F^{\mu\nu}) = 0, \nonumber \\
   \nabla_{\mu} \nabla^{\mu} \phi^{1/2} - \frac{1}{4} \phi^{3/2} F_{\mu\nu}F^{\mu\nu} = 0, \\ 
   R_{\mu\nu} = \frac{1}{2\phi^{1/2}} \left[ 2(\nabla_{\mu}\nabla_{\nu}\phi^{1/2} + \frac{1}{2} g_{\mu\nu} \square \phi^{1/2}) 
              + \phi^{3/2} ( F_{\mu\lambda}F^{\lambda}_{\nu} - g_{\mu\nu} F_{\alpha\beta}F^{\alpha\beta}) \right]. \nonumber  
\end{eqnarray}

In order to see if the Kaluza-Klein reduction of the Gibbons-Hawking surface term in equation (1) still leads to the appearance of the Abelian Chern-Simons term with the choice of the alternative parametrization given in equation (11), we follow exactly the same procedure as illustrated earlier, namely using $\hat{g}_{4} = \phi g_{3}$ with $\hat{g}_{4\hat{a}\hat{b}}$ and $g_{3ab}$ being metrics on $\partial M = \partial M_{4} \times S^{1}$ and $\partial M_{4}$ respectively and by assuming the "cylindricity" condition we can readily perform the dimensional reduction of the Gibbons-Hawking surface term. Further with the help of combined classical field equations of dilaton and photon on the 3-dimensional boundary $\partial M_{4}$, 
\begin{eqnarray}
   \nabla_{a} \phi^{1/2} - \frac{1}{2} \phi^{3/2} A^{b} F_{ab} = 0, \nonumber
\end{eqnarray}
the on-shell equivalent action to the dimensionally-reduced surface term takes the form: 
\begin{eqnarray}
   \int_{\partial M_{4} \times S^{1}} d^{4}\hat{x} \sqrt{\hat{g_4}} \hat{K} 
    = 2\pi R \int_{\partial{M_4}} d^3x \sqrt{g_3} \phi^{1/2} \left[ K + \frac{1}{4} \phi \epsilon^{abc} A_{a} F_{bc} \right]                                
\end{eqnarray}
where again the second term on the right hand side of equation (14) is precisely the Abelian Chern-Simons term defined on the 3-dimensional boundary of the 4-dimensional manifold $M_{4}$.

Therefore, the choice of an alternative parametrization in equation (1) still leads to the Einstein-Maxwell-Chern-Simons theory with dilaton described by the action:
\begin{eqnarray}
   S &=& \frac{1}{2\kappa^2} \int_{M_4} d^{4}x \sqrt{g} \phi^{1/2} R + \frac{1}{\kappa^2} \int_{\partial{M_4}} d^{3}x \sqrt{g_3} \phi^{1/2} K  \nonumber  \\  
     &+& \frac{1}{2\kappa^2} \int_{M_4} d^{4}x \sqrt{g} \phi^{1/2} \left(-\frac{1}{4} \phi F_{\mu\nu}F^{\mu\nu}\right)  \\
     &+& \frac{1}{2\kappa^2} \int_{\partial{M_4}} d^{3}x \sqrt{g_3} \phi^{1/2} \left(\frac{1}{2} \phi \epsilon^{abc}A_{a}F_{bc} \right).  
     \nonumber                     
\end{eqnarray}

Having obtained the dimensionally-reduced complete action, on a 4-dimensional manifold and its 3-dimensional boundary (which is not spacelike), now we would like to comment on its symmetries. As we already mentioned earlier, in the absence of the Gibbons-Hawking surface term, the reduced 4-dimensional action in equation (4) is invariant under the general coordinate transformations, the local U(1) gauge transformations, and the global scale transformations. 

First, we pointed out earlier that the classical field equations remain unchanged even if the Kaluza-Klein reduction of the Gibbons-Hawking surface term is added in the action. This fact obviously implies that both actions in quations (10) and (15) (corresponding to two alternative choices of the parametrization) still possess invariances under general coordinate transformations. 

The other two symmetries, i.e., the global scale invariance, namely, the dilatation symmetry and the local gauge invariance are broken by the addition of the Gibbons-Hawking surface term, though. In the absence of the surface term, both the action in equation (10) and the action in equation (15) used to be invarianct under the local U(1) transformations, $A_{\mu} \rightarrow A_{\mu} + \partial_{\mu} \Lambda$. 

But they are no longer invariant when the surface term, more concretely, the Chern-Simons actions are included. Likewise, without the surface term, the action in equation (10) is invariant under the global scale transformations of fields: $\phi \rightarrow \phi + \frac{2}{\sqrt{3}} \lambda \quad \& \quad A_{\mu} \rightarrow e^{\lambda} A_{\mu}$. 

And the action in equation (15) is invariant under the dilatations, $g_{\mu\nu} \rightarrow \lambda^{-1} g_{\mu\nu}, \quad \phi \rightarrow \lambda^{2} \phi \quad \& \quad A_{\mu} \rightarrow \lambda^{-3/2} A_{\mu}$.

Again, the addition of the surface term, the Chern-Simons action explicitly breaks these global scale transformations. These observations seem to teach us a lesson regarding the significant role played by the Chern-Simons term exclusively on a 3-dimensional spacetime. It has been generally acknowledged in the original version of Kaluza-Klein theory that the masslessness of the graviton is due to general covariance; the masslessness of the photon is due to gauge invariance and the masslessness of the dilaton is due to its being a Goldstone boson associated with the spontaneous breakdown of the dilatation. Thus the breakdown of the local U(1) gauge symmetry due to the Chern-Simons action in our extended version of the Kaluza-Klein theory indicates that the photon becomes "massive" when we consider its dynamics in a 3-dimensional spacetime. In other words, in the context of our extended version of the Kaluza-Klein theory, photons are "massive" and hence mediate "finite-ranged" electromagnetic interaction on a plane as opposed to what happens in the space of 3-dimensions. This kind of effect of the Abelian Chern-Simons term on dynamics on the plane indeed has been generally speculated and accepted in several currently flourishing topics such as anyon, superconductivity, and fractional quantum hall effect \cite{WIL}.

\begin{center}
{\rm\bf IV. DISCUSSION}
\end{center}

What has been shown in the present work is that, when properly extended by adding the gravitational surface term, the Kaluza-Klein theory is actually the higher dimensional unification of Einstein-Maxwell-Chern-Simons theory in the presence of a non-trivial scalar dilaton field. 

And the finite-ranged electromagnetic interaction due to massive photons on a plane is just one of the natural predictions of this extended Kaluza-Klein theory. Besides, in most of the studies in anyon physics the inclusion of the Chern-Simons term made up/out of some fictitious gauge field is considered to be optional. However, if we are willing to buy this extended Kaluza-Klein theory, the inclusion of the Chern-Simons term made up/out of the same gauge field as that in the Maxwell term is compulsory to describe planar dynamics. To our Knowledge, the Chern-Simons term of this sort and not of mathematical origin is believed to arise in the effective gauge theory action in 3-dimensional spacetime as a result of integrating out Fermionic degree of freedom in the theory.

But our extended Kaluza-Klein formulation reveals that the Maxwell term and the Chern-Simons term emerge on equal footing as fundamental terms in the gauge action.  

Naturally one might be tempted to generalize the present Abelian set-up to non-Abelian formulation eventually for the Kaluza-Klein type unification of gravity (General Relativity) and Yang-Mills theory. Such a non-Abelian generalization is indeed under consideration by the present authors. After all, the Kaluza-Klein theory deserves continuing attention and further exploration.

\vspace*{1cm}


\end{document}